# A study on the surface states of a topological insulator: $Bi_2Se_3$

Wasim Raja Mondal, [1] and Swapan K Pati [1]

Theoretical Sciences Unit[1]

Jawaharlal Nehru Centre for Advanced Scientific Research, Bangalore, 560064, India.

email addresse:

Wasim Raja Mondal: wasimr.mondal@gmail.com

We have presented an introductory study on surface states of topological insulator, $Bi_2Se_3$ based on the first principle.

# Introduction

The recent discovery of topological insulators opens up a new direction in condensed-matter physics [1, 2]. Topological insulator is a new class of material, in which its bulk is insulating while its surface states are conducting. It is different from ordinary semiconductor in the sense that surface states of topological insulator are robust against any external perturbation because these surface states are protected by time-reversal symmetry [3-5]. The response of surface states to any external perturbations bears great importance for the device applications in spintronics and to discover the new physics at the surface states of topological insulators.

Many binary and ternary chalcogenide compounds, for example, $Sb_2Te_3$ and $Bi_2Se_3$ have been discovered and predicted to be topological insulators [6-12]. Among these, $Bi_2Se_3$ is expected to be most promising candidate for its application. Its bulk band gap value is quite large and it can go up to 0.3 eV which is higher than the room temperature. Its band inversion occurs at Γ point, providing a simple band structure of the topological surface states with only single Dirac cone. The single Dirac cone can provide us a way to demonstrate the Berry phase which is originating from the spin-momentum locking pattern of the topological surface states that obey time-reversal symmetry. Many exciting phenomena, like, Majorana fermion, topological magneto-electric effects, quantized anomalous Hall effect, have been predicted. All these phenomena need opening of a gap at the Dirac point of the topological surface states. Although the gapless dispersion of the surface states is protected by the time reversal-symmetry, opening up of a gap violate the time reversal-symmetry. One of the easiest ways to break this time reversal symmetry is the application of magnetic doping which can open up a gap opening at the Dirac point [13-15]. So it is expected that surface perturbation can induce a band bending [16, 17]. This kind of band bending can anchor two-dimensional electron gas systems [16], which could show Rashba spin splitting [18]. Already, many studies have been devoted to test the surface states of $Bi_2Se_3$ and verified that the nontrivial surface state is particularly robust against different kinds of adsorbates [19-21]. But, in device application, the most important test is the response of the topological insulator surface state in ambient conditions. It has been already shown that oxygen can induce a p type doping [22], but the effect of moisture in the air or water vapor still is in mystery. To fulfill this gap in the surface state study of $Bi_2Se_3$, Benia et al [23] have come up with an experimental study describing the effect of water vapor on the surface states of $Bi_2Se_3$. Motivated by this experiment, in this chapter we have addressed the following question to explore the new physics of surface states of $Bi_2Se_3$: How and why do the water molecules interact with Se surface?

# Computational details

All calculations have been carried out using plane-wave based density functional theory (DFT) method as implemented in the PWSCF package [62]. The interactions between core electrons and valance electrons are approximated by ultrasoft pesudopotentials [63]. The exchange-correlation interaction of electrons is considered within the generalized gradient approximation (GGA) of Perdew-Burke-Ernzerhof (PBE) type [64] . Plane-wave basis sets are used with an energy cut off 30 Ryd for wave functions and a cut off 300 Ryd for the charge density. For the integration over the Brillouin zone, we consider 3× 3×1 Monkhorst Pack Mesh [64].The atomic relaxation is carried out by energy minimization using forces on each of the atoms in the

Broyden-Flecher-Goldfarb-Shanno (BFGS) method. Convergence of structure optimizations is obtained considering total energy difference between two consecutive steps less than $10^{-8}$ Ryd, the maximum force allowed on each atom is less than $10^{-3}$ Ryd/Bohr.

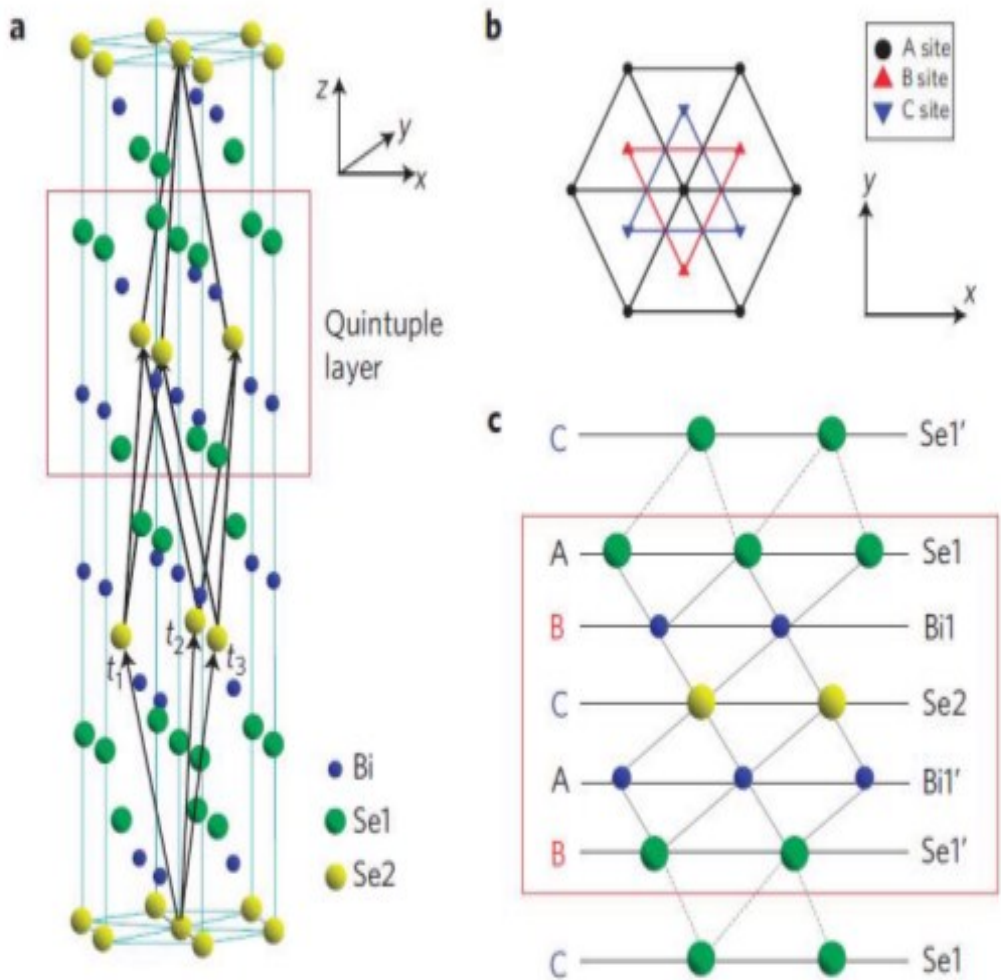

Figure 1. Crystal structure of $Bi_2Se_3$ [12]

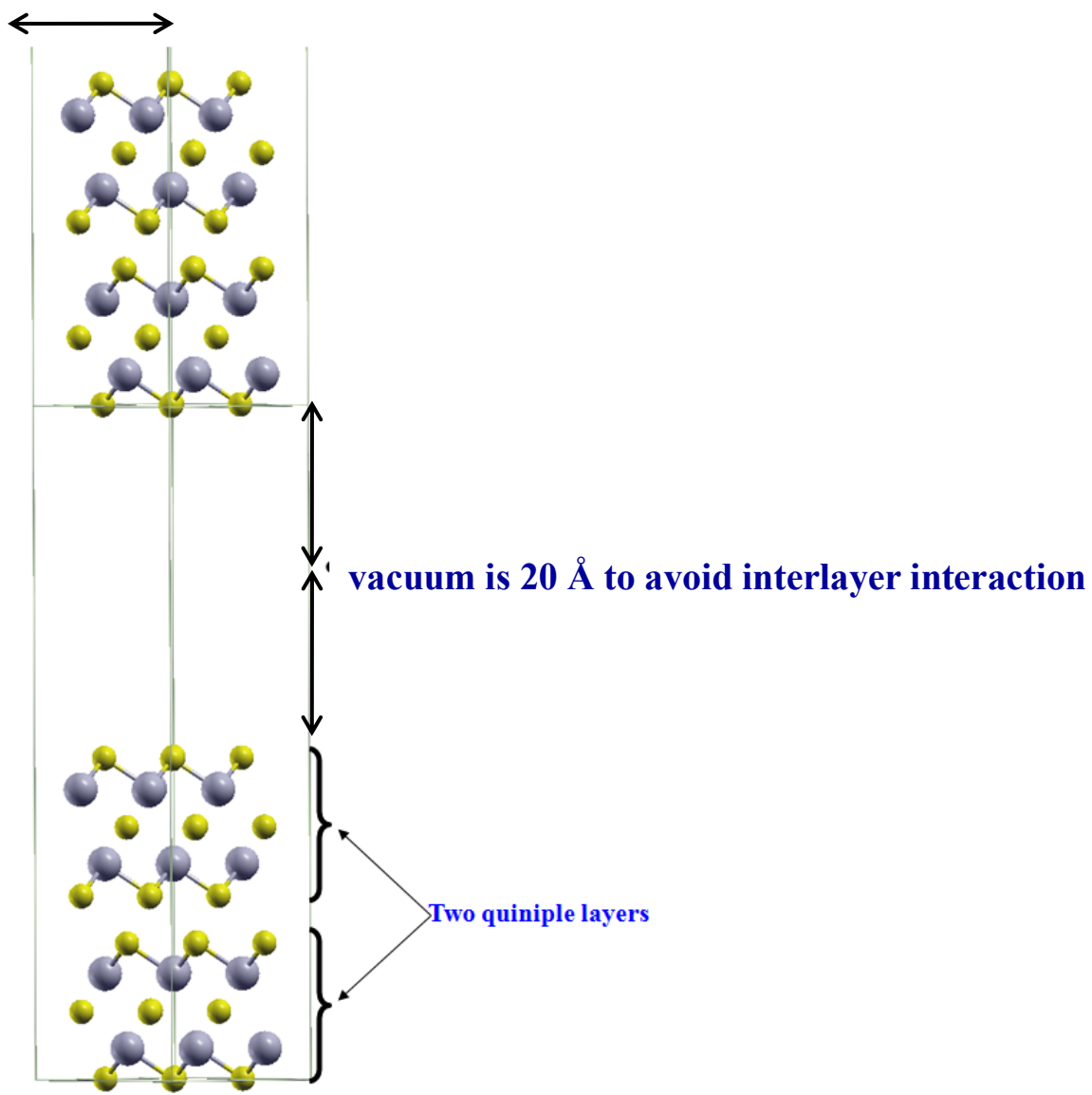

**Figure 2. Slab geometry of $Bi_2Se_3$**

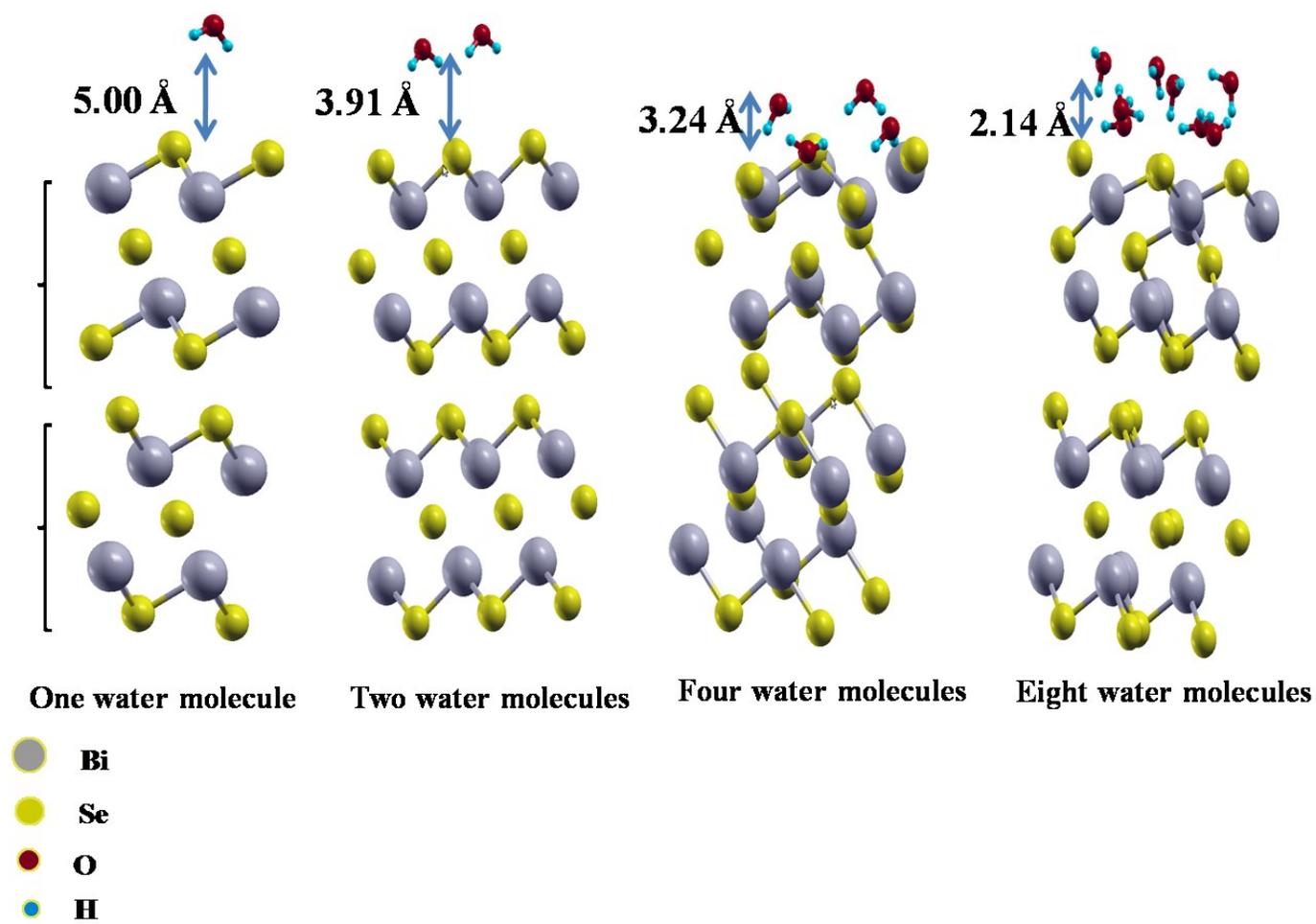

Figure 3. Water adsorption on the surface of $Bi_2Se_3$

# Results and Discussion

The unit cell of $Bi_2Se_3$ is rhombohedral with space group $D_{3d}^5(R\vec{3}m)$ with five atoms in one unit cell. It has layered structure and there is a triangular lattice within each layer (see Fig. 4.1). Z axis can be defined as trigonal axis which has three-fold rotation symmetry, X axis is a binary axis with two fold rotational symmetry and Y axis is bisectrix axis which is in the reflection plane. There are five atoms layers arranged along Z direction. These layers are called quintuple layers. There are five atoms with two equivalent Se atoms; two equivalent Bi atoms and a third Se atom in each quintuple layer (see Fig. 4.1 (c)). There is strong coupling between two atomic layers within one quintuple layer but there exists much weaker couling (van der Waals type) between two quintuple layers. The primitive lattice vectors ($t_1$, $t_2$, $t_3$) are shown in Fig. 4.1 (a). The Se2 site (see Fig. 4.1 (c)) plays the role of inversion centre. We have used 2×2×1 supercell with two quintuple layers containing 40 atoms for the simulation. The vacuum layer in the supercell is considered as 20 Å to avoid the artificial interlayer interactions (see Fig. 4.2). According to the experimental observations, in the presence of water, $Bi_2Se_3$ surface is modified through some reaction. To study this effect, we consider surface adsorption of water on the $Bi_2Se_3$ surface. We start with adsorption of one water molecule on the $Bi_2Se_3$ surface. We test for all the possible sites for the water adsorption. We find that the most preferable binding site is in between two Se atoms rather than on top of the Se atoms on topmost $Bi_2Se_3$ surface. For the water adsorption in this preferred binding site, we consider two, four and eight water molecules. The height of water molecule from the $Bi_2Se_3$ surface is found to be 5.00 Å, 3.91 Å, 3.24 Å and 2.14 Å for one, two, four and eight water molecules respectively (see Fig. 4.3).

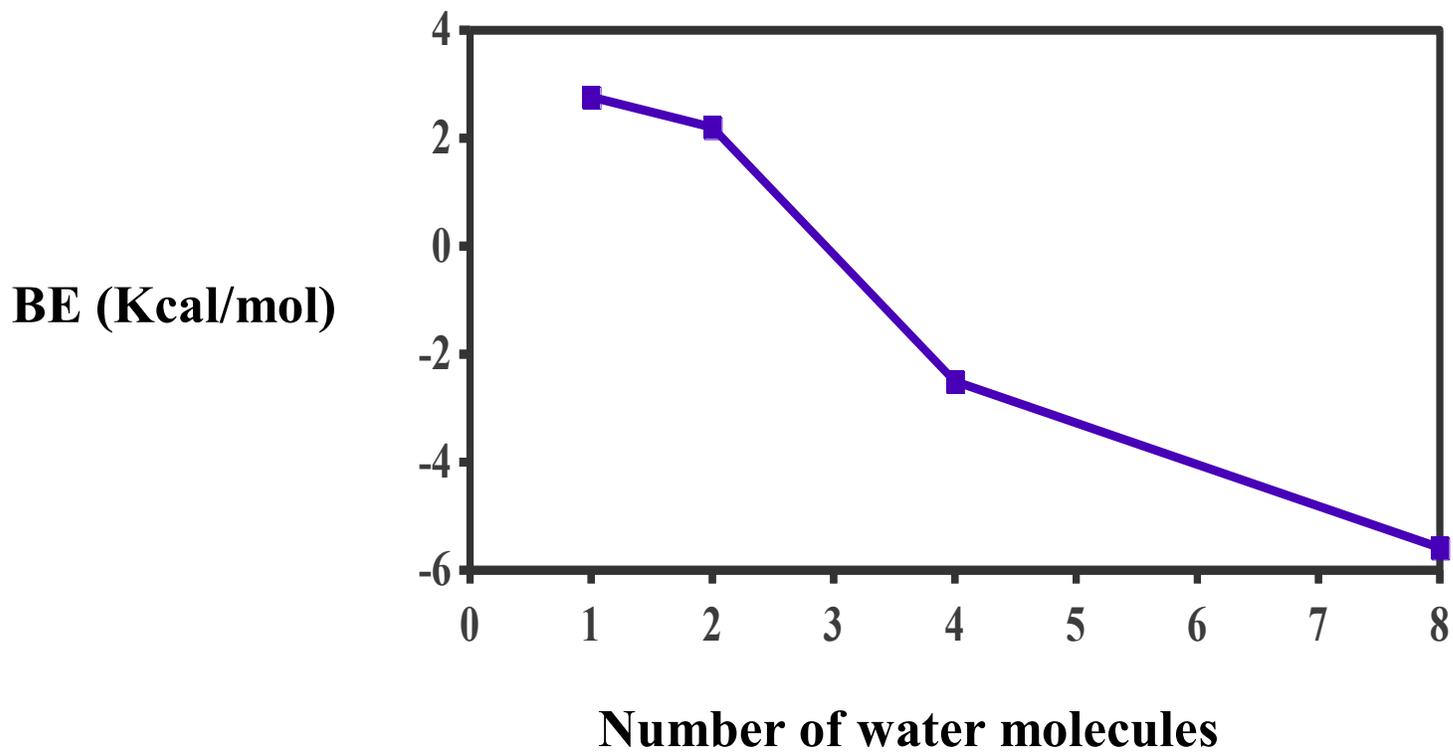

Figure 4. Plot of binding energy for different number of water molecules.

This manifests a low sticking coefficient for water which is consistent with experimental observation [23]. Experimentally it was found that after exposure to several hundred Langmuir of water, there is little change in the surface band structure. It is interesting to note that the intermolecular distance between the water molecules is comparable with hydrogen bond distance. So the hydrogen bond is important for the reaction of water with $Bi_2Se_3$ surface. The binding energies are calculated with the formula given by

$$BE = E_{Bi_2Se_3.nH_2o} - [E_{Bi_2Se_3} + nH_2o]$$

Here BE is the binding energy; $E_{Bi2Se3.nH2o}$ is energy of $Bi_2Se_3$ and $H_2o$ together, $E_{Bi2Se3}$ is the energy of $Bi_2Se_3$ and $nH_20$ is the energy of the n number of water molecules.

Fig. 4.14 shows that our calculated binding energy for various number of water molecule. We see that as the number of water molecules increase, the binding energy becomes more negative, favoring interactions stabilizations. We find that as the amount of water molecules increases, the water adsorption on the $Bi_2Se_3$ surface becomes more favorable.

## Conclusions

We have presented an introductory study on the surface of topological insulator, $Bi_2Se_3$. Our study shows that the reaction between surface states of $Bi_2Se_3$ and water is favorable when the surface is exposed to huge amount of water. This finding is consistent with experimental observations. We find that hydrogen bonding is important for this reaction.